# Maritime Design: A CSCW Territory?

Yushan Pan and Hans Petter Hildre

Norwegian University of Science and Technology, Ålesund 6025, Norway
{yushan.pan; hans.p.hildre}@ntnu.no

**Abstract.** This paper focuses on remote-control and autonomous vessels from a sociological perspective. We report that if CSCW research aims to shed light on other disciplines, researchers should be reflexive insider that first position themselves in such disciplines. Through reflexive practice, CSCW researchers could connect communities of practice, thus narrowing the distance between design and engineering.

## 1   Introduction

The literature shows that current maritime technology does not purely support cooperative work among operators on board [1]. The current design of operator–vessel interaction follows the principles of engineering design, including cognitive ergonomics and human factors [2]. The fundamental principle is to focus on the design applicability, the scope of the technical process, and the system structures to support the efficacy of machine use [3]. Operators are subjects in experimental work conducted to verify that a design is successful. However, the social aspects of human–vessel interaction have been largely dismissed. Moreover, operators are not encouraged to articulate their requirements, and the system design team is composed of a variety of specialists acting in the capacity of consultants to the project.

If the above are the facts, then how could CSCW researchers contribute to the design of maritime technology? There are a number of contributions from human factor field. However they are not convincible due to its organisational egocentrism regarding safety [4] and risk management [5]. The maritime domain needs an approach that can lead designers to support cooperative work. Thus, this work-in-progress paper provides some thoughts around this.

The short paper is organised as follows: In section 2, we present the state-of-art of maritime remote control. After that, in section 3 we present the design work in maritime domain. We reflect how communities of practices could be done in the maritime domain in section 4. The working-in-progress will be present in Section 5.

## 2   Designing Remote Control Systems in a Nutshell

In the current maritime research, remote control is considered to involve technical interactions with computer systems as well as social interactions among operators. Although designing remote control systems for autonomous vessels has led to debates between academia and the shipping industry [6], most autonomous technologies still only operate well in the situation for which engineering designers designed and programmed them through mainly simulation. The same also applies to automation [7]. Moreover, human intervention is still needed to handle complex situations [8] in addressing the remote control of autonomous vessels, sensor technology and its cyber security issues [19–23], digital twin [26] [31], and the so-called guidelines for autonomous shipping. Information about remote systems design is rapidly increasing on Google. A priority of engineering designers is to determine how available technologies, such as communication and networking technologies, could be designed to operate in maritime environments efficiently and effectively. The common sense that underpins these previous studies is the assumption that the systems will be well-designed to support human tasks, such as drawing patterns, creating models, and making sense of a machine's actions. Through a well-structured technology-centred experiment, as in most engineering design work, engineering designers expect that human factor specialists [14, 15] could investigate whether or not interfaces could be built to satisfy the operators. If so, what kinds of "human error" could be investigated?

## 3    The development of maritime design

Traditionally, maritime technology is designed in the fields of mechanical engineering, electrical engineering, electronic engineering, and even computer engineering. In these fields, the focus is on control systems, machinery, and the automation of maritime vehicles of any kind. The design process is purposeful, systemic, and iterative. Engineering designers conduct their work in various constraint conditions to find possible solutions for problems that are usually limited to the given scenarios. Engineering designers communicate with a small group of users, for whom the design follows a positivist paradigm with the intention to test a system. Design requirements are usually based on three principles: corporate, technology, and social [16]. The primary principle is that the corporation needs to generate design requirements in line with the company's organisational structures, strategic vision, and available resources, based mainly on the knowledge and expertise of the engineering designers. This principle does not change until social aspects challenge the company's frame through markets.

The work practices of operators are omitted because understanding operators' in-situ work practices is considered a technique used in requirements elicitation as well as cost effective and even has been limited and *ad hoc* for systems design [17]. This phenomenon is not surprising to CSCW researchers. However, it is surprising that when software plays an important role in the development of mechatronic systems in the maritime domain, only a few advantages of software design contributed by CSCW research are adopted by engineering designers in the maritime field. Although CSCW design now can successfully address the complexity, dynamics and uncertainty of work practices in real life, the design process is traditionally divided into software, electronics and mechanics, and every discipline emphasises its own approach to designing maritime technology. Furthermore, vocabularies and methodologies are divergent, which makes the collaboration between the disciplines greatly challenging.

## 4      Connecting of communities

The members of new generations of CSCW researchers know about human-centred computing, we know how to do fieldwork, and we even know how to translate our findings into special formats to communicate with systems developers [1]. However, we miss long-term engagement and design sensitive analysis in dealing with our reflections on how we connect different communities. Most CSCW research is iterative enough of its design process and does not challenge the lack of voices of confessional reflection [18] in our community. When researchers seek intervention as a bridge between research and practice, we might fall into our existing cognitive knowledge and create our own artificial worlds and seek our own language in doing design. We focus on exploring the inner symbolic space of a paradigm, and we try to convince others to believe that our languages are universal and useful [19]. We feel it might be wrong. If we do not accept procedure-oriented engineering design, is it correct to assume that CSCW can provide a solution? Suchman [20] suggested that we might need to find a customised solution rather than a universal solution. The challenge of this idea is not only the cognitive aspect of engineering design and CSCW research. It requires the development of radically new forms of scientific inquiry.

When we read the CSCW literature, it is a challenge. We assume that even though new generations hold two sets of knowledge—CSCW and software engineering—, we should have different perspectives on what we have read, and we should consider them equal contributions to our knowledge. However, this inner attribute of researchers becomes both we and others. Because CSCW researchers are not a designer of remote control systems and most work still depends on control engineering principles, inquiry requires extensive empirical data and practical concerns as well as a theoretical framework that might be perceived as disconnected from social construction [21]. We need to give my peers the tools to criticise our accounts of our work practice in the workplace. We also need to engineering designers the tools to investigate the usefulness of our contribution to them. By connecting community, we make it possible for the engineering designers to discuss the situation and to switch from a cooperative project where everyone had his own spot to engage in truly collaborative work. Moreover, both the engineering designers and researchers recognised the value of reflectivity even though it might differ among us. The engineering designers will find a way forward to be comfortable with the various interests and reflected on them in a dialogue to find a solution.

## 5      Future work

In this essay, we report challenges and opportunities for CSCW researchers in the engineering-oriented fields. In order for making contributions, CSCW researchers need to take the responsibility to help engineering designers understanding the social aspects of engineering design. Our work is in the early stage and we hope to engage more engineering design projects to practice insights from CSCW to shape the design policy, principles and practices in various fields, in particular the unstructured maritime field.